\documentclass[useAMS,usenatbib]{mn2e}
\usepackage[dvips]{graphicx}
\usepackage{amssymb}
\usepackage{amsmath}
\usepackage{rotating}
\usepackage{lscape}
\usepackage{pifont}
\usepackage{placeins}

\title[LLE for massive protostars]{Locally linear embedding: dimension reduction of massive protostellar spectra}
 \author[J.L. Ward, S.L. Lumsden]{J. L. Ward$^{1,2}$\thanks{E-mail:
j.l.ward@keele.ac.uk (JLW); s.l.lumsden@leeds.ac.uk (SLL)} and S. L. Lumsden$^{2}$\footnotemark[1]\\
$^{1}$Astrophysics group, Lennard-Jones Building, Keele University, Keele, ST5 5BG, UK\\
$^{2}$School of Physics and Astronomy, E. C. Stoner Building, University of Leeds, Leeds, LS2 9JT, UK}

\makeatother

\begin{document}

\date{Accepted 2016 June 21. Received 2016 June 20; in original form 2016 January 26}

\pagerange{\pageref{firstpage}--\pageref{lastpage}} \pubyear{2002}

\maketitle

\label{firstpage}

\begin{abstract}
We present the results of the application of locally linear embedding (LLE) to reduce the dimensionality of dereddened and continuum subtracted
near-infrared spectra using a combination of models and real spectra of massive protostars selected from the Red MSX Source survey 
database. A brief comparison is also made with two other dimension reduction techniques; Principal Component Analysis (PCA) and Isomap using
the same set of spectra as well as a more advanced form of LLE, Hessian locally linear embedding.
We find that whilst LLE certainly has its limitations, it significantly outperforms both PCA and Isomap in classification of 
spectra based on the presence/absence of emission lines and provides a valuable tool for classification and analysis of large spectral data sets.
\end{abstract}


\section{Introduction}

Advancements in instrumentation and observing facilities have lead to an ever increasing rate of production of spectroscopic data. It 
is therefore an increasingly complex task to extract relevant data from large data sets. Dimension reduction algorithms offer the 
ability to reduce complex data sets into the lowest number of free parameters necessary to address a specified degree of variation.
The most commonly used dimension reduction algorithm applied to astronomical data is principal component analysis (PCA; \citealt{Deeming1964,Jolliffe1986})
however it is most sensitive to globally linear variations and has proven inefficient in classification based on varying emission line strengths \citep{Yip2004}.

Locally linear Embedding (LLE) is a manifold mapping dimension reduction algorithm introduced by \cite{Roweis2000}. It has already been 
successfully applied to the classification of galaxies and QSOs \citep{Vanderplas2009} and the separation of stars, galaxies and quasars
\citep{Daniel2011}, both using data from the 
Sloan Digital Sky Survey (SDSS; \citealt{York2000}). It has more recently been applied by \citet{Bu2013} to the classification of stellar
subclasses for M-type stars.
In these previous studies it was shown to provide useful distinction between classes of object based on features in the spectra
which relate to real physical properties.
In all previous applications LLE has proven to reduce the dimensionality of data more efficiently than principal component analysis (PCA)
and is better at separating spectral classes.

LLE attempts to compute low-dimensional embeddings of high-dimensional input data whilst preserving the local neighbourhood of each
input vector. The basis of the technique is to map an underlying lower-dimensional manifold upon which the high-dimensional data lies. When 
applying this technique to spectra, each spectrum is treated as a vector, $\vec{X_{i}}$, with a number of dimensions equal to the number
of wavelength bins of the spectrum. For example a spectrum with a wavelength range of 1.5  $\mu$m to 2.5  $\mu$m where each element
covers 13.351 \AA{} is treated as a 750 dimensional vector. We summarise the LLE algorithm in the following steps:
\begin{itemize}
 \item  Taking a set of n vectors $\vec{X_{i}}$ with dimensionality D, the k nearest neighbours for each vector are calculated based on 
   Euclidean distance. These k nearest neighbours form the local neighbourhood of $\Vec{X_{i}}$.
\item Linear weights to each of the nearest neighbours are calculated. This is done by minimising the cost function:
      \begin{equation}
     \epsilon(W) = \sum_{i}\vert \vec{X_{i}} - \sum_{j=1}^k W_{ij} \vec{X_{j}}\vert^{2}
    \end{equation}
\item        The inner products between each vector and each of its nearest neighbours are computed
    to produce a neighbourhood correlation matrix, $C_{jk} = \vec{X_{j}} \cdot \vec{X_{k}}$.
\item  The reconstruction weights are then computed using:
        \begin{equation}
     W_{ij} = \sum_{k}C-{jk}^{-1}(\vec{X_{i}}\cdot\vec{X}_{k}+\lambda)
    \end{equation}
    To ensure that the embedding is only based on a vectors
    local neighbourhood, $W_{ij}$ must be equal to 0 if $X_{j}$ is not one of the nearest neighbours of $X_{i}$.
    The sum of all of the weights
    to the neighbours of a single vector, $\lambda$, is equal to 1 (i.e. $\sum_{j} W_{ij} = 1$). This ensures that every  neighbourhood used in 
    the embedding is equally valid and therefore no single data point can distort the final embedding. This is done using the 
    Lagrange multiplier, $\lambda = \alpha/\beta$, where $\alpha = 1-\sum_{jk}C_{jk}^{-1}(\vec{X_{i}}\cdot\vec{X_{k}})$ and $\beta = \sum_{jk}C_{jk}^{-1}$.
\item Finally the embedded vectors $\vec{Y_{i}}$ which will make up the final output data are calculated by minimising a second cost function:
    \begin{equation}
     \Phi(Y) = \sum_{i}\vert\vec{Y_{i}}-\sum_{j}W_{ij}\vec{Y_{j}}\vert^{2}
    \end{equation}
where the weights $W_{ij}$ are known and the set of vectors $\vec{Y_{i}}$ need to be calculated. The constraint that $\sum_{i}\vec{Y_{i}} = \vec{0}$
is imposed in order to centre the projection on the origin and a second constraint is imposed in order to avoid degenerate solutions;
    \begin{equation}
     \frac{1}{n}\sum_{i}\vec{Y_{i}}\otimes\vec{Y_{i}} = I
    \end{equation}
    where I is the d$\times$d identity matrix (d being the dimensionality of the projected vectors).
\item     This cost function now defines a quadratic form containing the N$\times$N symmetric matrix $M_{ij}$:
    \begin{equation}
     \Phi(I) =\sum_{ij}M_{ij}(\vec{Y_{i}}\cdot\vec{Y_{i}})
    \end{equation}
    where $M_{ij} = \delta_{ij}-W_{ij}-W_{ij}+\sum_{k}W_{ki}W_{kj}$ and $\delta_{ij} = 1$ if $i = j$ and $\delta_{ij} = 0$ otherwise.
\item The lowest d+1 eigenvectors of the matrix $M_{ij}$ are computed with the exception of the bottom eigenvector (which represents a 
free translation mode of eigenvalue 0) in order to find the optimum embedding.
The embedded output is a representation of the variance between input data and is completely independent of the original form of the data.
\end{itemize}

The Red MSX\footnote[1]{Midcourse Space Experiment \citep{Egan2003}} Source (RMS) Survey \citep{Lumsden2013} is the largest catalogue of Massive Young Stellar Objects (MYSOs) and Ultra
Compact H\,{\sc ii} (UCH{\sc ii}) regions to date.
This survey has led to a series of follow up observations including near infrared spectroscopy of the largest collection of MYSO candidate objects to
date using the UK InfraRed Telescope (UKIRT; \citealt{Cooper2013a}) and the New Technology Telescope (NTT; Ward et al., in preparation). 
It is from this set that the spectra used in this
work have been selected.

In this paper, we apply the LLE algorithm to available \textit{HK} spectra of Galactic MYSOs in order to assess the effectiveness of 
dimension reduction algorithms as a method of automated classification.
In section 2 the input data will be described and the methodology used will be outlined. The results will be presented in section
3 while section 4 will provide a comparison with other dimension reduction algorithms for the same dataset.

\section{Input data and methodology}

The majority of \textit{HK} spectra of MYSOs can be separated into 4 distinct types based on the different emission line features that are present
and it is believed that these  types likely represent an evolutionary sequence.
The types and the rational behind the types, are explained in detail in Lumsden et al. (in preparation) but the 
main emission line classification criteria of the types are outlined in table 1.

  \begin{table}
  \caption{Formalisation of MYSO type classification criteria}
   \begin{tabular}{lllll}
\hline
    Type & H$_{2}$ & Br$\gamma$ & Br10 & $Fluor.$ Fe\,{\sc ii} \\
    \hline
    I & Present & Absent & Absent & Absent \\
    II & Present & Present & Absent & Absent \\
    III & Present & Present & Present & Present \\
    IV & Absent & Present & Present & Present \\
\hline
   \end{tabular}

  \end{table}

\subsection{Real Spectra}

The real reduced spectra used in this work are presented, along with investigations into the properties of the spectra,
in two other papers; \citet{Cooper2013a} and Ward et al., in preparation.
All spectra were reduced and prepared for input into the LLE algorithm using the \textsc{figaro} data reduction package and the
\textsc{dipso} spectral analysis package. The \textsc{snip} function of \textsc{dipso} was used to remove data in the range 17800 $< \lambda <$ 19700 \AA{}
as this region of the spectrum has the poorest atmospheric transmission. The spectra were then re-binned
into 750 elements in the range 15000--25000 \AA{} using the \textsc{scrunch} operation in \textsc{figaro}.

At this point the data is split into two sets; the dereddened spectra and the continuum subtracted spectra.
For the former, the spectra were dereddened using the \textsc{dered} function in \textsc{dipso} with extinction values from
\citet{Cooper2013a} and Ward et al., in preparation.
The continuum subtracted data had third order polynomials fitted to the continua which were then subtracted.

The final step in the preparation of the real spectra was to cross-correlate the spectra to a reference point. The reference spectrum used for 
all cross-correlation was that of the compact H\,{\sc ii} region G274, chosen for its prominent and extensive set of Brackett series emission lines.
Any spectra for which the absolute wavelength shift was greater than one element were discarded.
The remaining spectra were then classified into the types shown in Table 1, discarding any that did not fall into any of these categories or 
where the type was uncertain. This yielded a remaining 184 spectra (94 from \citealt{Cooper2013a} and 90 from Ward et al. in preparation), 
with 28 type I sources, 79 type II, 35 type III, and 42 type IV. The average real spectra for each type are shown in the upper four panels
of Fig. 1.
\begin{figure*}
 \begin{minipage}{175mm}
\includegraphics[width=0.9\linewidth]{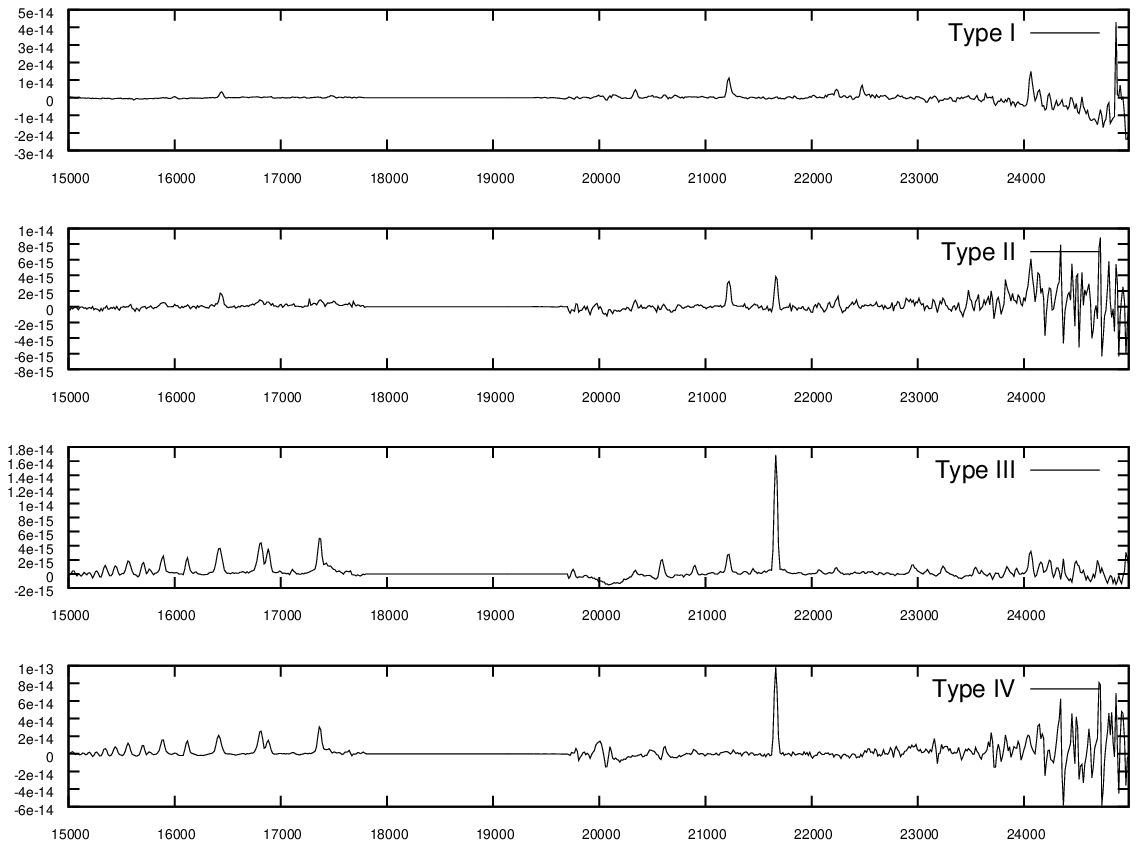} 
\includegraphics[width=0.9\linewidth]{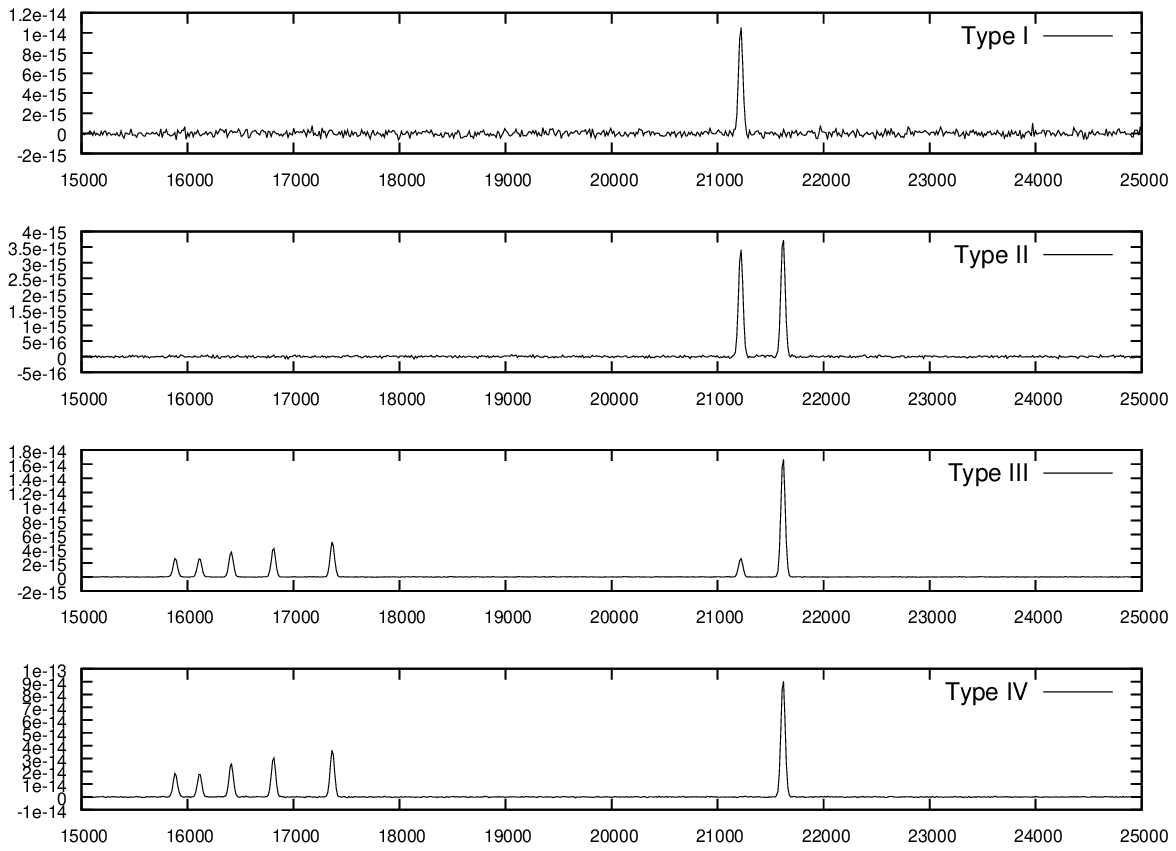} 
\caption{Upper: average spectra for each of the types I-IV calculated from continuum subtracted real spectra. Lower: average spectra for each of the types I-IV
calculated from the model spectra.}
 \end{minipage}
\end{figure*}

\subsection{Models}
Preliminary tests showed that the low number of spectra used in this study did not provide any significant result when 
the LLE algorithm is applied. 
We therefore include an additional large set of model spectra to act as a framework for the real spectra.
Two main sets of models have been created to complement the two sets of real spectra; a dereddened set of models and a continuum subtracted
set. To generate these models
a series of simple python scripts were used.

The continuum of the dereddened set of models was calculated using a blackbody distribution with a randomly generated temperature in the range 0 $<$ T $<$ 40000 K for
each spectrum and the wavelength at each element to calculate the appropriate flux. The values in the range 17800 \AA{} to 19700 \AA{}
were set to zero to match the section which was removed from the real spectra.The emission lines were then included through the addition
of Gaussian profiles using the rest wavelength of each emission line as a mean and an amplitude coefficient. The amplitude coefficients were
selected so that the mean peak fluxes of the model lines equal the mean peak fluxes of the real emission lines and are related to the 
peak flux by
\begin{equation}
  f_{peak} = A_{e} \cdot \frac{1}{\sigma_{l} \sqrt{2 \pi}} \cdot \frac{1}{2 \sigma_{l}^{2}} \cdot 10^{-24}
\end{equation}
where $f_{peak}$ is the peak flux, $A_{e}$ is the amplitude coefficient and $\sigma_{l}$ is the line width.
Finally randomly generated noise was added to each point along the spectrum using a Gaussian distribution with a mean set to zero and 
where the $\sigma$ is inversely proportional to $\lambda^{4}$ to simulate the increase in noise from the \textit{K}-band to the \textit{H}-band which 
is exaggerated by the dereddening process in real spectra. The high levels of noise in the long wavelength end of the \textit{K}-band is 
neglected but this area of the spectra will not be included in the inputs for the dimension reduction algorithms.
A total of 8000 models were generated for both the dereddened and continuum subtracted data sets, equally divided between the four types and with 
the average model spectra for each type shown in the lower four panels of Fig. 1.
 The average model spectra exhibit significantly lower levels 
of noise than the average real spectra because of the much larger number of models.
All of the models generated for this work were based on making random variations surrounding a set of mean line profiles
and a mean continuum (in the case of the dereddened models) which were designed to match those of the average real spectra.

\subsection{Methodology}

The LLE node \citep{Vanderplas2009} from the Modular Data Processing toolkit (MDP; \citealt{Zito2009}) was used to perform the Locally Linear embedding. A range of values for the 
number of nearest neighbours, k were used in order to find the optimum sampling for the data. This was to ensure that the quality of
the resulting projections was not limited by the number of nearest neighbours used.
The LLE node of MDP determines the output dimensionality required to express a specified percentage of the variance 
in a similar manner to that suggested by \citet{deRidder2002}.
First a covariance matrix is calculated for each local neighbourhood and an eigenanalysis is performed to find the minimum number of dimensions
needed to represent the specified percentage of local variance.
The output dimensionality used in the LLE projection is then the mean of the dimensionalities calculated for the local neighbourhoods \citep{Vanderplas2009}.

As mentioned earlier, the longer wavelength end of the \textit{K}-band was not used for Locally Linear Embedding with the cut-off point at
2.28 $\mu m$. This point was chosen in order to avoid the CO bandhead which, as it does not affect the classification of the spectrum,
would introduce an unnecessary complication into the dataset. It also avoids the high levels at noise at wavelengths greater than
2.35 $\mu m$ which again is irrelevant to the classification criteria set out in the previous section.
All spectra were normalised to a distance of 1 kpc prior to dimension reduction using distances from the RMS survey database.

\section{Results}

\subsection{Distinguishing MYSOs from standard stars}

\begin{figure*}
\begin{minipage}{175mm}
 \includegraphics[width=0.49\linewidth]{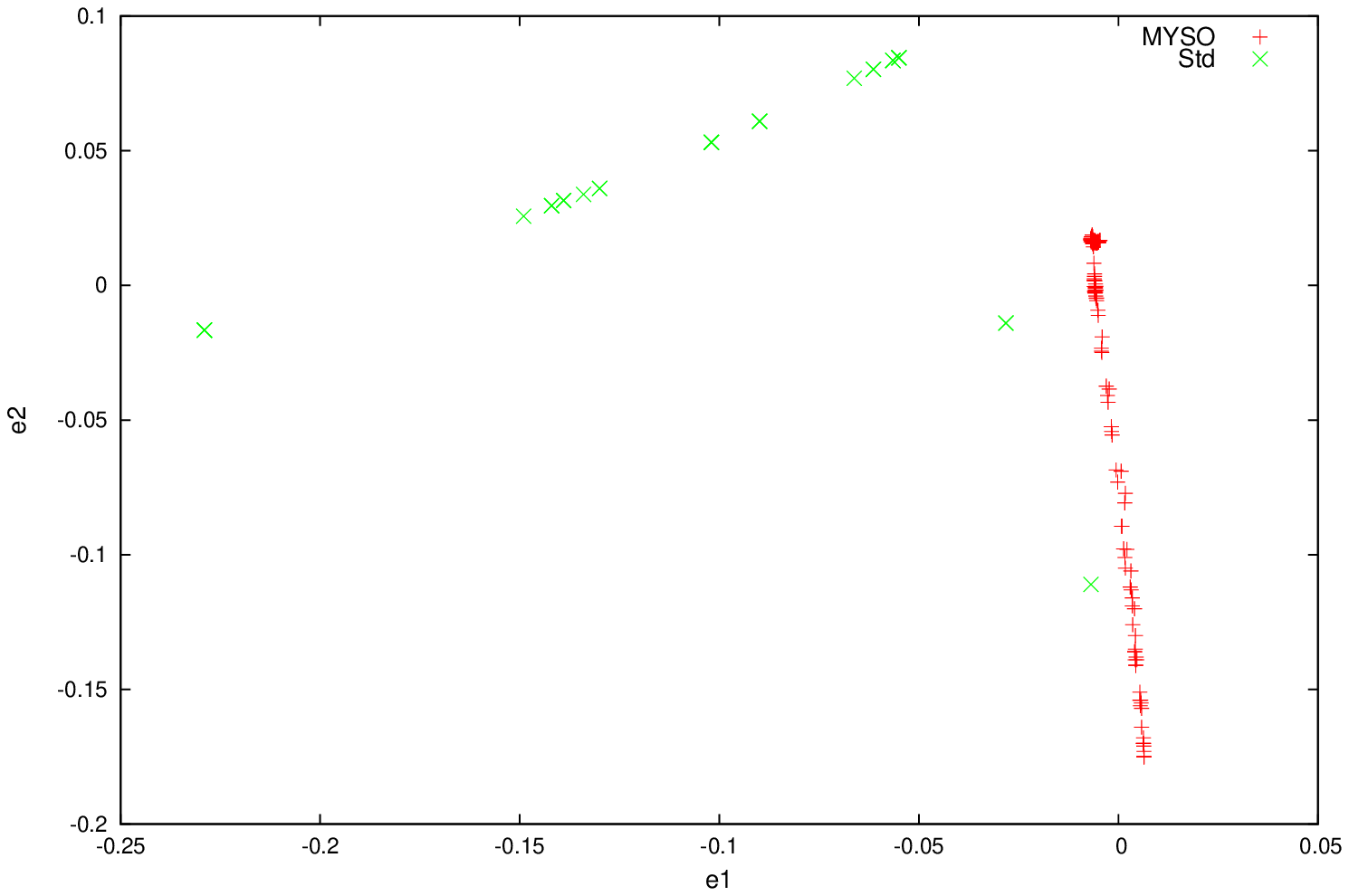}
  \includegraphics[width=0.49\linewidth]{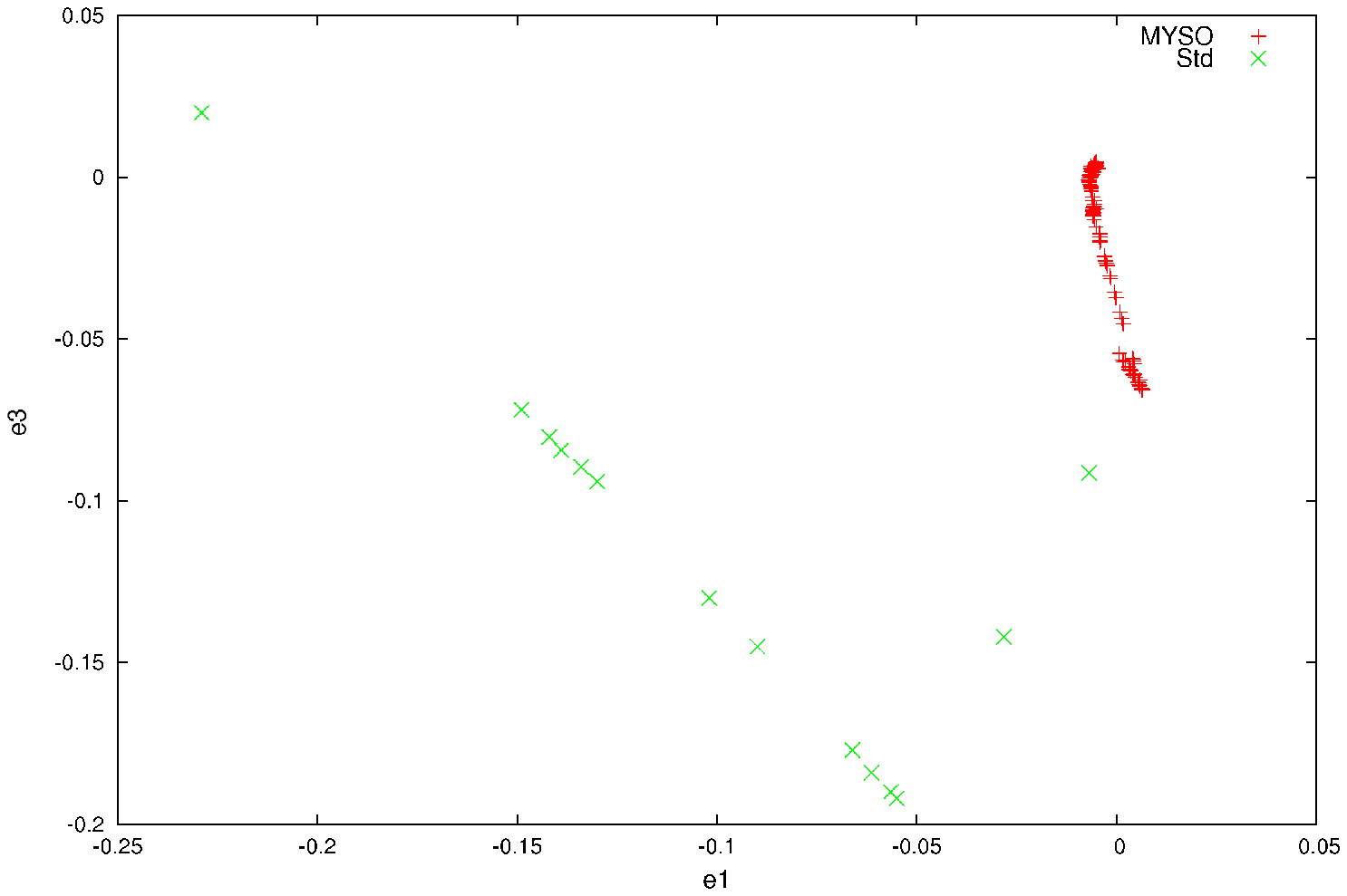}
\caption{Left: output e2 and e1 values of LLE applied to real MYSO spectra along with the telluric standard star spectra
using nearest neighbours, k$=$8.
Right: output e3 vs e1 values of LLE applied to real MYSO spectra with telluric standard spectra.}
\end{minipage}
\end{figure*}
As a preliminary test we used reduced and normalised spectra (before dereddening) of both
MYSOs and the telluric standard calibration stars from the same observing runs. 
Prior to the application of LLE to these data, both the standard star spectra and the MYSO spectra were 
re-binned using {\sc scrunch} as in the preparation procedure for the main LLE runs. They were then normalised by the 
mean value of the continuum emission for each spectrum. This ensures that the projection will be independent of 
total NIR continuum brightness.
Fig. 2 shows the 3 lowest eigenvector output dimensions of the LLE embedding and in both e2 vs e1 space and e3 vs e1 space the MYSO spectra
form a line which is clearly distinguishable from the standard spectra.
It is therefore apparent that LLE could provide a method for automated classification in larger spectroscopic surveys.

\subsection{Classification of MYSOs}

Preliminary tests using the 8000 model dereddened spectra found that the the spectra are indistinguishable based on type, forming 
simple polynomial shaped projections in the output eigenspace. This is attributed to the significant continuum emission in the 
spectra, which represents the dominant source of variation.  
It is apparent however that for a data set consisting of a large
number of less embedded spectra, LLE may provide an efficient method for estimating the temperatures of all of the spectra 
simultaneously as the temperature is the dominant source of variation between these models (see Fig. 3).

\begin{figure*}
\begin{minipage}{175mm}
\begin{center}
     \includegraphics[width=0.7\linewidth]{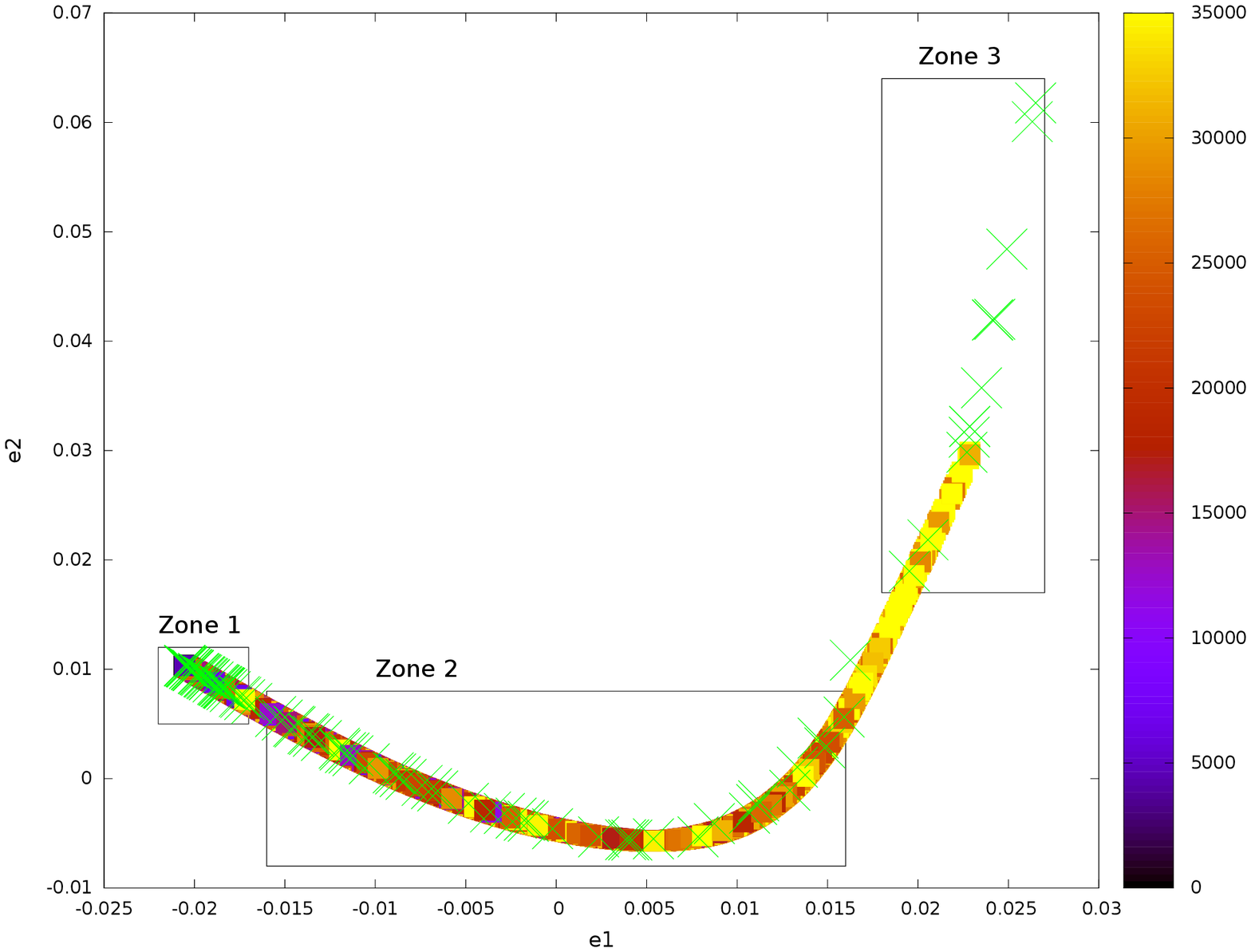}
     \includegraphics[width=0.7\linewidth]{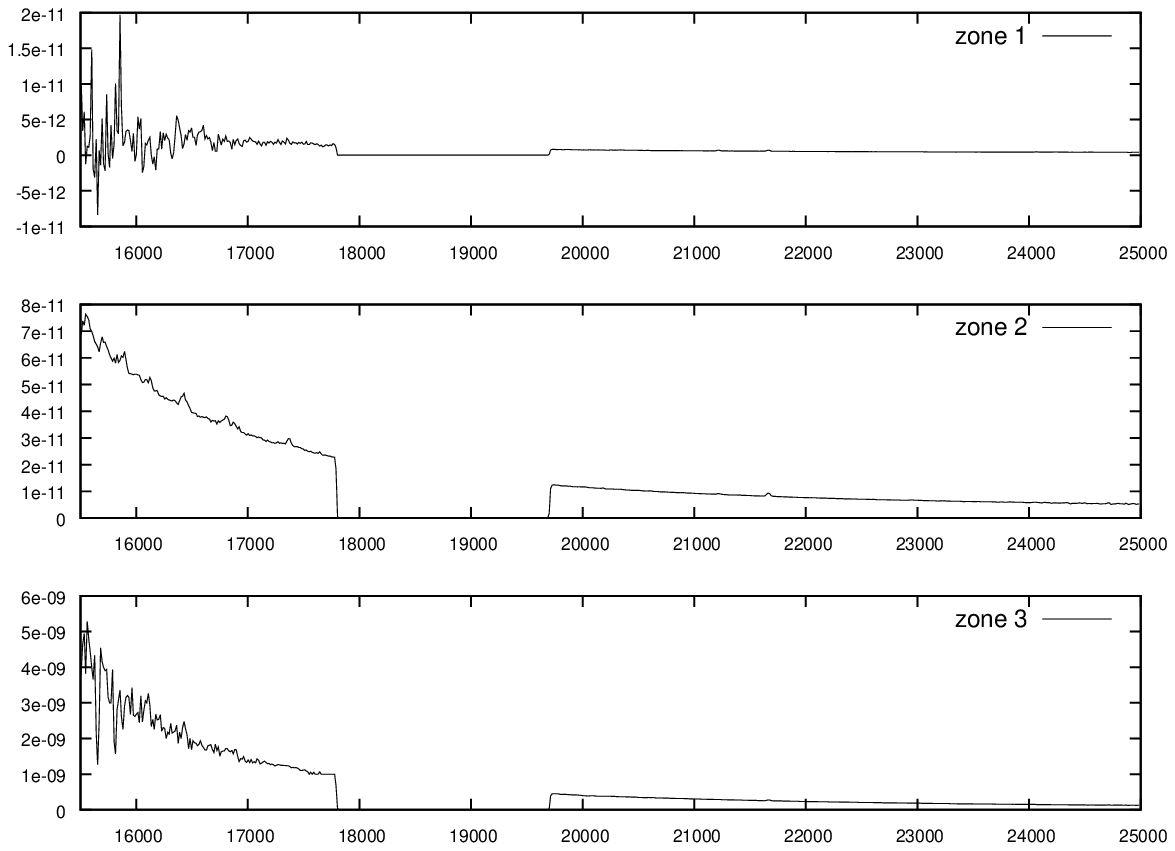}
\caption{Upper panel: LLE eigenvector outputs e2 vs e1 for model spectra, colour coded by model blackbody temperature and real spectra (green Xs). 
Lower panel: average spectrum for the real spectra in each of the zones in the left panel.}
\end{center}
\end{minipage}
\end{figure*}

As no distinction can be made between types based on raw or dereddened spectra due to the dominance of continuum emission, the next
logical step is to perform LLE on spectra without continuum emission.
The process of continuum subtraction based on a simple polynomial fitting
 is a step which can be easily automated for most spectral datasets and would therefore would be reasonable to include in an 
initial classification algorithm for a large spectroscopic survey.

Our preliminary tests using only continuum subtracted real spectra of MYSOs yielded no significant results that we were able to 
interpret. The most likely explanation for this is that there are not enough spectra included in order to produce a reasonable 
projection of the variation within the data based on any physical properties.

In order to account for 95\% of the variation between spectra in our data set only three output dimensions were required when using LLE.
The lowest two output dimensions (representing the largest variation) of the LLE projection when applied to our model and real continuum subtracted spectra are shown in Fig 4.
When LLE is applied to our model continuum subtracted spectra it is immediately clear that the separation of 
the model types is significantly improved compared with the application to dereddened spectra. The projection in e2 vs e1 resembles two main branches, one of 
type I and II spectra and one of type III and IV spectra with an area of mixing between type II and III spectra close to the point at which the two
main branches converge. This point of convergence coincides with the spectra with the lowest signal-to-noise ratios.

The real type I spectra coincide well with their model counterparts, as do the majority of the type II 
sources. There is considerable mixing between the type III/IV sources and some of the type II sources however. Regardless of any mixing however, a clear trend 
can be seen from positive e1 and negative e2 values in type I spectra and negative e1 with positive e2 values in types III and IV. 
With more robust models which accurately represent the Fe\,{\sc ii} emission and the complete Brackett series it is likely that 
this distinction will be clarified.

\begin{figure*}
 \begin{minipage}{175mm}
  \includegraphics[width=0.49\linewidth]{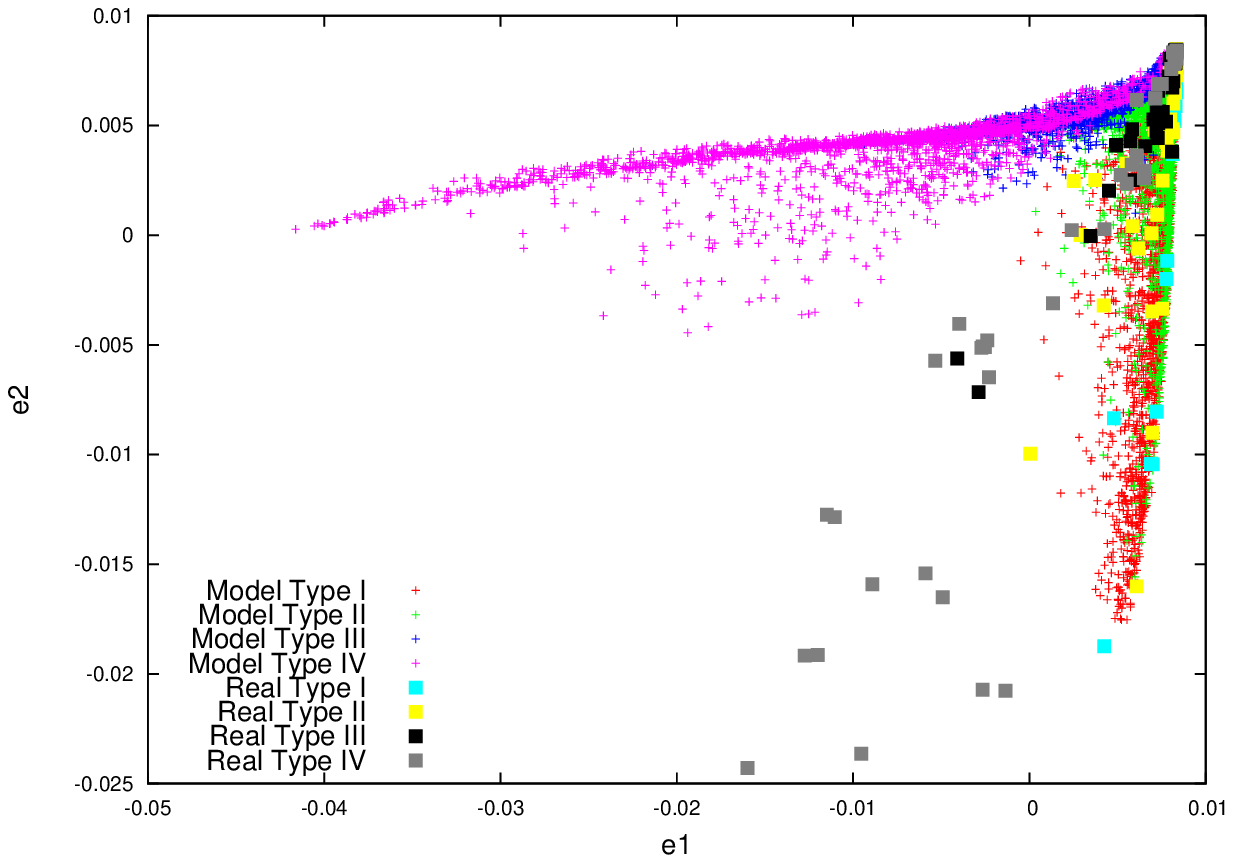}
  \includegraphics[width=0.49\linewidth]{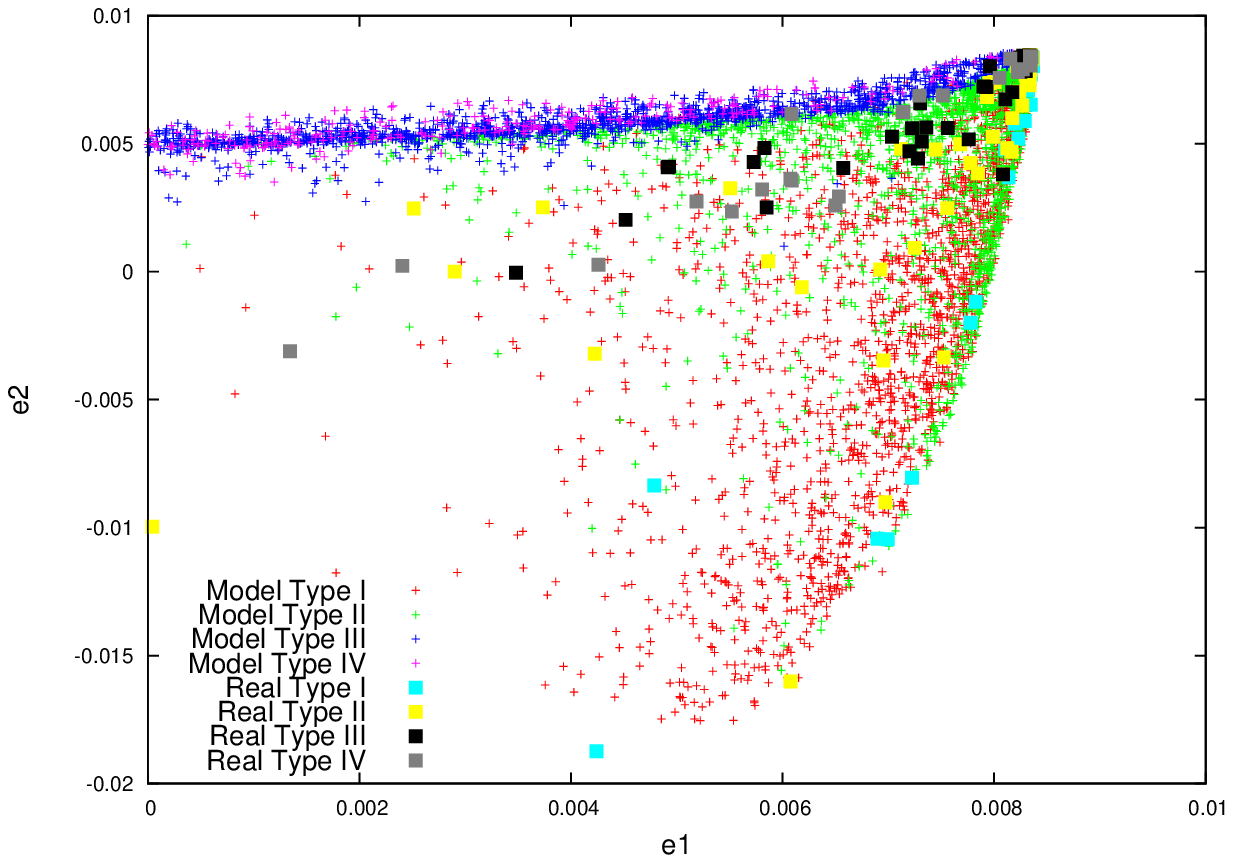}
\caption{Left: LLE e2 vs e1 output for models plus real spectra with nearest neighbours k=100. Right: same as left but showing only $0 < \text{e}1 < 0.01$.}
 \end{minipage}
\end{figure*}

A simple classification script has been written incorporating LLE which automatically assigns a type based on the location 
of the object in e2/e1 space using the angle and distance of the sources from the point at which the I/II and III/IV branches
converge.
We find that even with this simplistic approach 63\% of the real spectra were successfully determined as belonging to 
either types I/II or III/IV and for type I/II spectra, this figure rises to 79\%. Whilst this leaves many spectra without 
classification, less than 3\% of the spectra were actually mistakenly assigned incorrect types.

\section{Comparison with Other Dimension Reduction Algorithms}

We will now compare our outputs of the LLE dimension reduction algorithm with those of the similar, widely used dimension reduction algorithms
PCA and Isomap. Finally we will compare the output of a more advanced version of LLE; Hessian locally linear embedding
(HLLE).

\subsection{Principal component analysis and Isomap}

Principal component analysis requires 431 output dimensions in order to account for 95\% of the variance between our (model and real) spectra. As a 
dimension reduction algorithm therefore, it has proved to be inefficient compared with LLE.
Figure 5 shows the resulting second and first eigenvector outputs of the application of PCA to the same data set as LLE is applied to in Fig. 4.
It is immediately apparent that all of the spectra are clumped together.
We find similar effects in all output dimensions and conclude that for such similar spectra, PCA cannot be used to accurately distinguish between our different types in this way.
\begin{figure}
 \includegraphics[width=\linewidth]{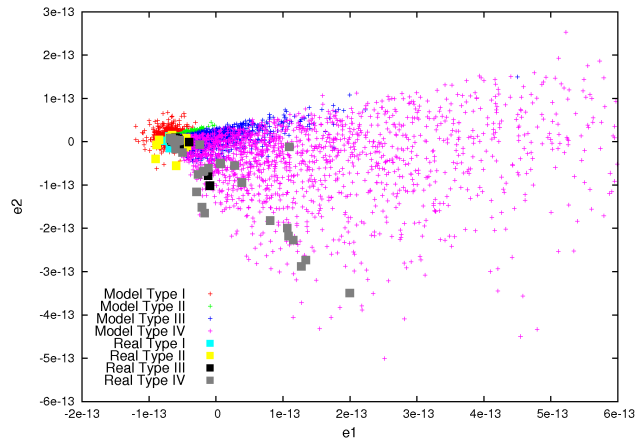}
\caption{PCA output with dimensions equivalent to those output from LLE. The types of the data points correspond directly to those of Fig. 4.}
\end{figure}
Performing Isomap \citep{Tenenbaum2000} on our set of model spectra (the output is shown in Fig. 6) we find that a type I and type IV spectra are more 
adequately separated than when PCA is applied. The Isomap output does however lack much of the finer structure of the LLE output as a
result of losing much of the local neighbourhood information which is preserved in the LLE algorithm.
\begin{figure}
 \includegraphics[width=\linewidth]{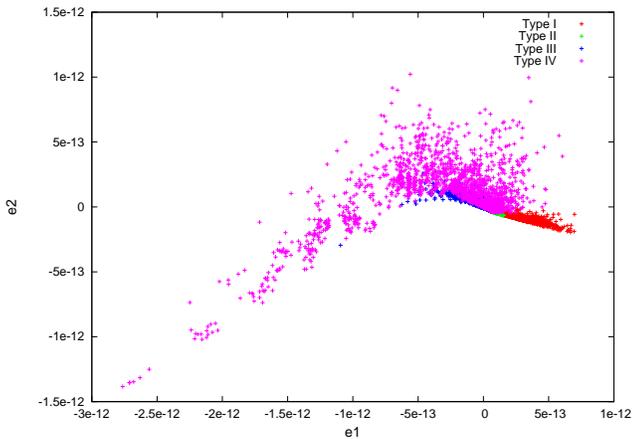}
\caption{Isomap output with dimensions equivalent to those output from LLE.}
\end{figure}

\subsection{Hessian locally linear embedding}

Finally we apply the HLLE algorithm \citep{Donoho2003} to the same data to see whether our results with LLE can be improved upon.
HLLE differs from the standard LLE algorithm through the use of Hessian estimators in place of the weights in the original LLE
procedure. 
The nearest neighbours are found by calculating Euclidean distances and the matrix $M$ is produced for each neighbourhood as in LLE.
Then a singular value decomposition of $M$ is performed to obtain tangent coordinates and the Hessian estimator is developed;
  \begin{equation}
   H_{i,j}=\sum_{l}\sum_{r}(H_{r,i}^{l}H_{r,j}^{l}) \text{,}
  \end{equation}
where $H^{l}$ is a $d(d+1)/2\times k$ matrix associated with estimated the Hessian over the neighbourhood N$_{i}$
The eigenvectors corresponding to the lowest eigenvalues are computed from the matrix $H$ similar to the LLE analysis of the matrix $M$. 
Embedding coordinates are obtained from the matrix $ W = V^{T}R^{-1/2}$ where $R=VV^{T}$.

The output embedding e2 vs e1 plot for the HLLE algorithm is shown in Fig. 7. When compared to the equivalent output from the standard LLE algorithm 
in Fig. 4 we find that the application of HLLE gives no marked improvement over that of LLE for this particular data set. This is most likely
due to the continuous nature of the model production which provides a data set that is inherently complete without any gaps.
It is likely therefore that the HLLE algorithm would provide a greater advantage where no model spectra are available.
\begin{figure}
 \includegraphics[width=\linewidth]{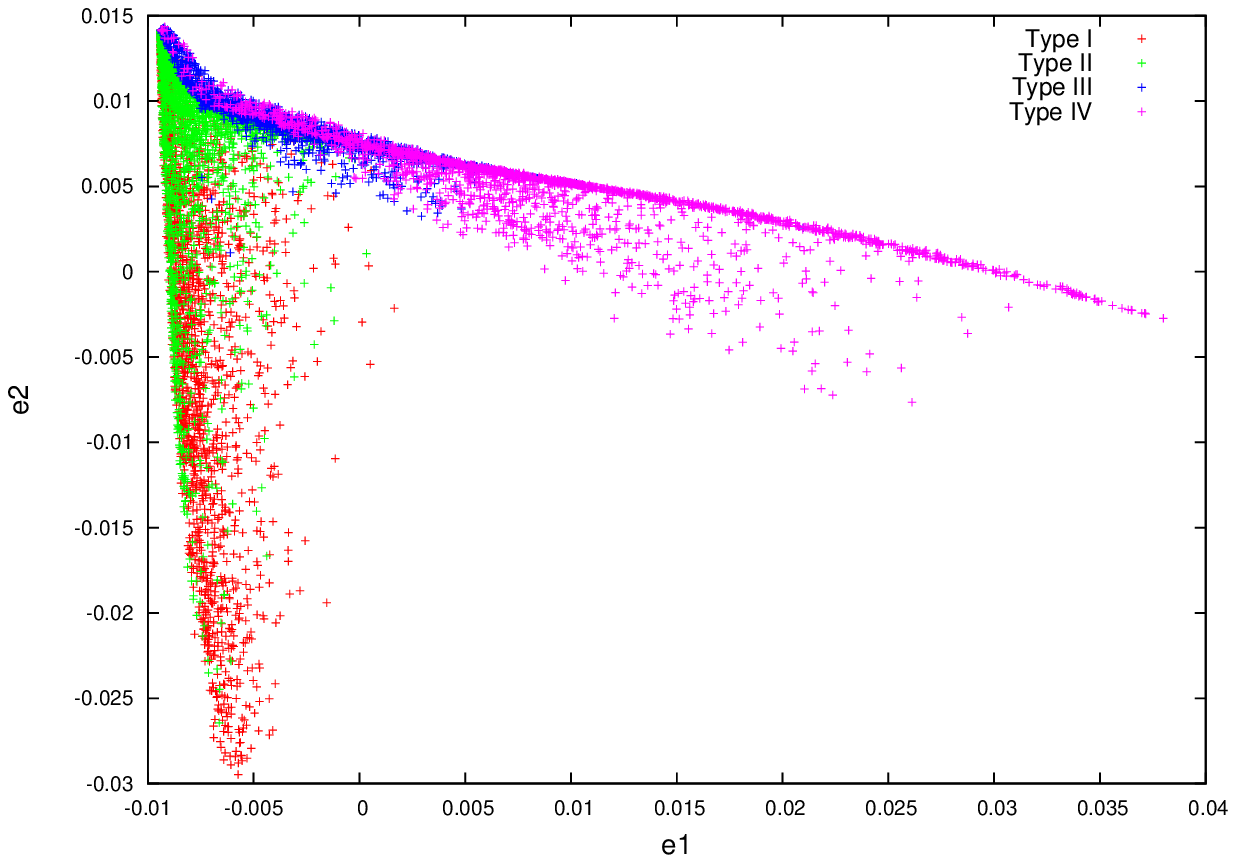}
\caption{HLLE output with dimensions equivalent to those output from LLE.}
\end{figure}

\section{Conclusions}

We have successfully applied LLE to a relatively small set of real spectra using a framework of 8000 model spectra.
The conclusions of this application of LLE to massive protostellar spectra are summarised below:
\begin{enumerate}
 \item LLE has successfully been utilised to separate MYSO spectra from telluric standard stars with no training sets or modification to the spectra
 and it is therefore likely that LLE would be particularly useful in the early analysis of large spectroscopic surveys to classify sources.
\item It was not possible to separate the dereddened spectra based on their emission lines because they are 
dominated by continuum emission. It may be possible in some cases however to determine properties of the continua of spectra
and for objects without significant continuum emission (relative to the line emission/absorption) this would not be an issue.
\item Continuum subtracted spectra were separated into the expected types using LLE and through a simple automation of the 
interpretation of the output data almost two thirds of the spectra were successfully assigned a type whilst only 2 sources were 
misclassified.
\item LLE has shown itself to be superior to PCA and Isomap in terms of the efficiency of dimension reduction and separation
of MYSO spectra based on the presence and absence of relevant emission lines. Little improvement was found when replacing LLE 
with HLLE but this would likely be more effective with a less continuous set of data than our models.
\end{enumerate}

Whilst computationally more expensive than many of the available alternative dimension reduction algorithms, LLE and its derivatives
HLLE and robust LLE present a powerful tool for classifying large spectroscopic datasets, outperforming the more commonly utilised PCA and Isomap algorithms.
With the advent of the next generation of observing facilities and ever more advanced multi-object spectrographs, the 
ability to quickly analyse and classify large samples of spectra will play a fundamental role in future large scale surveys.

\section*{Acknowledgments}
The authors thank the anonymous referee for his/her useful comments.
JLW acknowledges financial support from the Science and Technology Facilities Council of the UK (STFC) via the PhD studentship programme.
This paper made use of information from the Red MSX Source survey database at http://rms.leeds.ac.uk/cgi-bin/public/RMS\_DATABASE.cgi 
which was constructed with support from STFC. 
This research has made use of the SIMBAD data base, operated at CDS, Strasbourg, France.

\bibliographystyle{mn2e}
\bibliography{LLE_bib}

\label{lastpage}

\end{document}